\documentclass[letter,longauth]{aa} 

\usepackage{graphicx}
\usepackage{lscape}
\usepackage{longtable}
\usepackage{natbib}
\usepackage{color}
\usepackage{array} 
\usepackage{tikz,array}
\usetikzlibrary{calc}
\usepackage{multirow}
\usepackage{here}
\usepackage{txfonts}
\usepackage{amsmath,amstext}
\usepackage{epstopdf}
\usepackage{float}
\usepackage{mathtools}
\usepackage{booktabs}
\usepackage{subfigure}
\usepackage{url}
\usepackage{helvet}
\usepackage{tabularx}
\usepackage{multirow}
\usepackage{natbib}
\usepackage[flushleft]{threeparttable}
\usepackage{lscape}
\usepackage{pdflscape}
\usepackage{longtable}
\usepackage{wasysym}
\usepackage{float}
\usepackage{relsize}
\usepackage{color}
\usepackage{placeins}
\usepackage{breqn}
\usepackage{bm}
\usepackage{enumitem}
\usepackage{makecell}

\usepackage[colorlinks, citecolor=blue, linkcolor=blue]{hyperref}
\hypersetup{colorlinks,breaklinks, linkcolor=blue,urlcolor=magenta, anchorcolor=blue,citecolor=blue}
\bibpunct{(}{)}{;}{a}{}{,} 

\newcommand{\be}{\begin{equation}}
\newcommand{\ee}{\end{equation}}

\def\gcm3{\hbox{g cm$^{-3}$}}       
\def\Msun{\hbox{$\mathrm{M}_{\astrosun}$}}             
\def\Rsun{\hbox{$\mathrm{R}_{\astrosun}$}}

\def\Mearth{\hbox{$\mathrm{M}_{\oplus}$}}
\def\Rearth{\hbox{$\mathrm{R}_{\oplus}$}}

\newcommand{\gaia}{\textit{Gaia}}
\newcommand{\kepler}{\textit{Kepler}}
\newcommand{\ktwo}{K2-399\,b}
\newcommand{\target}{EPIC\,248472140}
\newcommand{\targetA}{EPIC\,248472140\,A}

\usepackage{stfloats}

\usepackage{graphicx}
\usepackage{txfonts}
\begin{document}

   \title{K2-399\,b is not a planet}

   \subtitle{The Saturn that wandered through the Neptune desert is actually a hierarchical eclipsing binary\thanks{This work is based on observations collected \emph{a}) at the Centro Astron\'omico Hispano en Andaluc\'ia (CAHA) at Calar Alto, operated jointly by the Instituto de Astrof\'isica de Andaluc\'ia (CSIC) and the Junta de Andaluc\'ia; \emph{b}) at the European Southern Observatory (ESO) under ESO programmes 0100.C-0808 and 108.21YY; \emph{c}) at Roque de los Muchachos Observatory with the Italian Telescopio Nazionale \textit{Galileo} (TNG) operated by the INAF - Fundaci\'on Galileo Galilei, under the OPTICON program 2017B/059.}}

   \author{
J.~Lillo-Box\inst{\ref{cab}}, 
D.~W.~Latham\inst{\ref{cfa}}, %
K.~A.~Collins\inst{\ref{cfa}},  
D.~J.~Armstrong\inst{\ref{warwick},\ref{ceh}},
D.~Gandolfi\inst{\ref{torino}},
E.~L.~N.~Jensen\inst{\ref{swarthmore}},
A.~Castro-Gonz\'alez\inst{\ref{cab}},
O.~Balsalobre-Ruza\inst{\ref{cab}},
B.~Montesinos\inst{\ref{cab}},
S.~G.~Sousa\inst{\ref{ia-porto}},
J.~Aceituno\inst{\ref{caha}},  
R.~P.~Schwarz\inst{\ref{cfa}},   
N.~Narita\inst{\ref{komaba},\ref{osawa},\ref{iac}}, 
A.~Fukui\inst{\ref{komaba},\ref{iac}},
J.~Cabrera\inst{\ref{uberlin}},
A.~Hadjigeorghiou\inst{\ref{warwick}},
M.~Kuzuhara\inst{\ref{osawa},\ref{japan-oan}},
T.~Hirano\inst{\ref{osawa},\ref{japan-oan}},
M.~Fridlund\inst{\ref{onsala},\ref{leiden}},\\
A.~P.~Hatzes\inst{\ref{tautenburg}},
O.~Barrag\'an\inst{\ref{oxford}},
N.~M.~Batalha\inst{\ref{ucsc}}
}

\titlerunning{K2-399\,b is not a planet: the Saturn that wandered through the desert is a hierarchical eclipsing binary}
\authorrunning{Lillo-Box et al.}

\institute{
Centro de Astrobiolog\'ia (CAB), CSIC-INTA, ESAC campus, Camino Bajo del Castillo s/n, 28692, Villanueva de la Ca\~nada (Madrid), Spain \email{Jorge.Lillo@cab.inta-csic.es} 
\label{cab}
\and
Center for Astrophysics \textbar \ Harvard \& Smithsonian, 60 Garden Street, Cambridge, MA 02138, USA 
\label{cfa}
\and
Department of Physics, University of Warwick, Gibbet Hill Road, Coventry, UK 
\label{warwick}
\and
Center for Exoplanets and Habitability, University of Warwick, Gibbet Hill Road, Coventry, UK 
\label{ceh}
\and
Dipartimento di Fisica, Universit\`a degli Studi di Torino, Via Pietro Giuria, 1, 10125, Torino, Italy \label{torino}
\and
 Dept. of Physics \& Astronomy, Swarthmore College, Swarthmore, PA 19081, USA
 \label{swarthmore}
\and
Instituto de Astrof\'isica e Ci\^encias do Espa\c{c}o, Universidade do Porto, CAUP, Rua das Estrelas, 4150-762 Porto, Portugal
\label{ia-porto}
\and
Centro Astron\'omico Hispano en Andaluc\'ia, Sierra de los Filabres sn, 04550, G\'ergal Almer\'ia, Spain 
\label{caha}
\and
Komaba Institute for Science, The University of Tokyo, 3-8-1 Komaba, Meguro, Tokyo 153-8902, Japan\label{komaba}
\and
Astrobiology Center, 2-21-1 Osawa, Mitaka, Tokyo 181-8588, Japan 
\label{osawa}
\and
Instituto de Astrof\'isica de Canarias (IAC), 38205 La Laguna, Tenerife, Spain 
\label{iac}
\and
Institut f\"ur Planetenforschung, Deusches Zentrum f\"ur Luft- und Raumfahrt, Rutherfordstr. 2, 12489 Berlin, Germany \label{uberlin}
\and
National Astronomical Observatory of Japan, 2-21-1 Osawa, Mitaka, Tokyo 181-8588, Japan \label{japan-oan}
\and
Department of Space, Earth and Environment, Chalmers University of Technology, Onsala Space Observatory, 43992, Sweden \label{onsala}
\and
Leiden Observatory, University of Leiden, PO Box 9513, 2300 RA, Leiden, The Netherlands \label{leiden}
\and 
Th\"uringer Landessternwarte Sternwarte 5, D-07778,  Tautenburg, Germany \label{tautenburg}
\and
Sub-department of Astrophysics, Department of Physics, University of Oxford, Oxford, OX1 3RH, UK \label{oxford}
\and
Department of Astronomy and Astrophysics, University of California, Santa Cruz, CA 95060, USA
\label{ucsc}
}

   \date{In press.}

 
  \abstract
   {The transit technique has been very efficient in detecting planet candidate signals over the past decades. The so-called statistical validation approach has become a popular way of verifying a candidate's planetary nature. However, the incomplete consideration of false positive scenarios and data quality can lead to the misinterpretation of the results.}
   {In this work we revise the planetary status of K2-399\,b, a validated planet with an estimated false positive probability of 0.078\% located in the middle of the so-called Neptunian desert, and hence a potential key target for atmospheric prospects.}
   {We use radial velocity data from the CARMENES, HARPS and TRES spectrographs, as well as ground-based multi-band transit photometry LCOGT MuSCAT3 and broad band photometry to test the planetary scenario.}
   {Our analysis of the available data does not support the existence of this (otherwise key) planet, and instead points to a scenario composed of an early G-dwarf orbited {in a $846.62^{+0.22}_{-0.28}$~days period} by a pair of eclipsing M-dwarfs (hence a hierarchical eclipsing binary) likely in the mid-type domain. We thus demote K2-399\,b as a planet.}
   {We conclude that the validation process, while very useful to prioritise follow-up efforts, must always be conducted with careful attention to data quality while ensuring that all possible scenarios have been properly tested to get reliable results. We also encourage developers of validation algorithms to  ensure the accuracy of a priori probabilities for different stellar scenarios that can lead to this kind of false validation. We further encourage the use of follow-up observations when possible (such as radial velocity and/or multi-band light curves) to confirm the planetary nature of detected transiting signals rather than only relying on validation tools.}


   \maketitle
%
\section{Introduction}

After the launch of the \textit{Kepler} mission \citep{borucki10}, the first Earth-size and sub-Earth-size planets were detected with the transit technique \citep[e.g., ][]{batalha11,barclay13,sanchis-ojeda13}.  In contrast to gas giants, the radial velocity signature of the small planets detected by \textit{Kepler} was out of reach of high-precision ultra-stable spectrographs owing to the faintness of most \textit{Kepler} host stars (V > 13), which made radial velocity (RV) follow-up observations difficult. In this context, the validation process was proposed \citep{torres11,fressin11,morton11}. The validation approach consists of the statistical rejection of alternative non-planetary scenarios that could reproduce the observed transit signal. Based on the transit properties (e.g., duration and shape), and optionally fed by ancillary observations such as low-precision radial velocity data or high-spatial resolution imaging, the proposed algorithms can determine an overall probability that the transit signal comes from another non-planetary scenario, known as false positive probability (FPP). {Different authors have established in the past different thresholds to consider a transit signal as a validated planet (e.g., \citealt{torres15}, \citealt{rowe14,morton16,armstrong21,mantovan22,castro-gonzalez22}).} 

Depending on the different catalogs, this validated disposition is directly put at the same level as the so-called confirmation process (typically involving a mass measurement for the planet). As an example, the NASA Exoplanet Archive\footnote{\url{https://exoplanetarchive.ipac.caltech.edu}.} \citep{akeson13} or the \texttt{exoplanets.eu}\footnote{\url{https://exoplanet.eu/home/}.} \citep{schneider11} catalogs, being the two most relevant in the field, do not distinguish among these two different dispositions. This may have key implications in population synthesis studies and several other follow-up observing programs. In this context, an exoplanet confirmation protocol {needs} to be discussed among the exoplanet community to agree on the principles that qualify a planet detection as confirmed.
{Revising the validation process and establishing a confirmation protocol will be of especial relevance in the context of the new PLATO mission \citep{rauer14} and forthcoming new facilities such as the ELT, the Habitable Worlds Observatory (HWO) or the proposed mission LIFE \citep{quanz21}.}

\ktwo{} was presented by \cite{zink21} as a planet candidate based on data from the repurposed version of the \kepler\ mission, K2 (\citealt{howell14}). The authors identified a transit signal with an ultra-short period of $\sim$0.76 days and a planet-to-star radius ratio of $\sim$0.035 (corresponding to $\sim 6$~\Rearth{} for the derived stellar parameters resulting in an F9 dwarf), with a grazing eclipse of impact parameter $\sim$0.97. According to the stellar properties derived in this discovery paper, the semi-major axis of the planet candidate orbit was only $\sim$1.9 times the stellar radius, becoming one of the closest planets to its parent star. More interestingly, its size and period made this planet belong to the so far unpopulated Saturn and Neptune desert \citep{benitez-llambay11, szabo11, youdin11}, a key region of the parameter space sculpted by formation and migration processes that is still under debate (e.g.,  \citealt{castro-gonzalez24}, in press.).

The planet candidate was subsequently validated by \cite{christiansen22} using the \texttt{vespa}\footnote{\url{https://github.com/timothydmorton/VESPA}.} code \citep{vespa,vespa-soft}. {The authors used high-resolution spectra and high-spatial resolution imaging to feed this algorithm, and obtained}
an FPP of $7.8\times10^{-4}$, thus validating the signal as coming from a planetary origin. It is interesting to note that the authors point out the high \gaia{} \citep{gaia} \texttt{ruwe}\footnote{Renormalised Unit Weight Error}
value of this star, being 5.89 (where values above 1.4 indicate a bad astrometric solution typically because of the presence of an unresolved star or a long-period sub-stellar companion). {However, their centroid motion analysis provided a high confidence of the transit occurring on-source.} 
Consequently, the authors considered the signal as statistically validated. As such, this validated planet appears with the "Confirmed Planet" disposition in the NASA Exoplanet Archive\footnote{\url{https://exoplanetarchive.ipac.caltech.edu/overview/K2-399}.} and the "Confirmed" planet status at the \texttt{exoplanets.eu}\footnote{\url{https://exoplanet.eu/catalog/k2_399_b--9074/}.} catalog. 

In this letter, we provide additional follow-up observations (presented in Sect.~\ref{sec:observations}), whose analysis in Sect.~\ref{sec:analysis} allows us to demote the planetary status of this signal and to propose a more likely alternative scenario in Sect.~\ref{sec:alternative}. In Sect.~\ref{sec:conclusions}, we conclude with some final remarks. 
Given the nature of this letter, we will use \ktwo{} when referring to the claimed planet, while the \target{} naming convention to refer to the system itself.

\section{Observations and stellar characterization}
\label{sec:observations}

\subsection{TRES spectroscopy}
NASA's Transiting Exoplanet Survey Satellite (TESS, \citealt{ricker14}) mission independently identified transit-like events in this star, named TIC\,374200604 in the TESS Input Catalog, \citealt{stassun19}) and released it as TESS Object of Interest (TOI) TOI-4838 in early 2022. This led almost immediately to follow-up reconnaissance spectroscopy with the Tillinghast Reflector Echelle Spectrograph (TRES\footnote{\url{http://www.sao.arizona.edu/FLWO/60/tres.html}.}, PI: A.~Szentgyorgyi) on the 1.5-m Tillinghast Reflector at the Fred Lawrence Whipple Observatory on Mount Hopkins (Arizona, USA). TRES is a fiber-fed CCD spectrograph with resolving power $R=44,000$ and wavelength coverage 384 to 909 nm. 
The six TRES observations obtained for this target cover a wide time span of 2275 days and show large RV variations at the several km/s level, with a median uncertainty per datapoint of 30 m/s. These RVs are listed in Table~\ref{tab:RVs}.

\subsection{HARPS and HARPS-N spectroscopy}
\label{sec:HARPS}

The host star candidate K2-399 was selected as one of the key targets of the NOMADS (PI D. Armstrong; see, e.g., \citealt{osborn23}) and KESPRINT (PI: D. Gandolfi; see, e.g., \citealt{gandolfi17}) observing programs with the HARPS\footnote{High Accuracy Radial velocity Planet Searcher.} instrument at the 3.6-m telescope at La Silla Observatory (Chile) of the European Southern Observatory (ESO), before the candidate planet was validated. 
A total of 12 spectra were obtained through the KESPRINT program\footnote{Program ID: 0100.C-0808.} between 23-February-2018 and 15-March-2018, while 27 measurements were acquired through the NOMADS program\footnote{Program ID: 108.21YY.} from the 31-January-2023 to 28-April-2023. The whole dataset was reduced by using the Data Reduction Software (DRS) version v3.8{, and absolute radial velocities were extracted by using the cross-correlation technique \citep{baranne96} with a G2 mask.}
From the cross-correlation function (CCF) the radial velocity, as well as the shape properties of the CCF (bisector span and FWHM) were obtained. The average RV uncertainty from this dataset is 7.0 m/s with a standard deviation of the uncertainties corresponding to 0.8 m/s. The RVs and activity and related indicators are shown in Table~\ref{tab:RVs}. Two additional spectra were secured\footnote{This was part of the KESPRINT observing program (Program ID: OPT17B_59 or 2017B/059) of K2 transiting planet candidates (PI: A.\,P.\,Hatzes; see, e.g., \citealt{prieto-arranz2018})} with the HARPS-N spectrograph \citep{cosentino12} mounted at the 3.58-m Telescopio Nazionale \textit{Galileo} (TNG) of Roque de los Muchachos Observatory (La Palma, Spain). The data reduction, as well as the extraction RVs follows the same procedures as for the HARPS spectra.

\subsection{CARMENES spectroscopy}
\label{sec:CARMENES}
We observed \target{} in two campaigns with the CARMENES instrument (\citealt{quirrenbach10}) installed at the 3.5\,m telescope at Calar Alto observatory. 
In the first campaign (PI: M.~Kuzuhara), we acquired 20 measurements in 4 nights between 27-November-2018 and 26-February-2019. In our second campaign (PI: J.~Lillo-Box) in 2024, on 7-January-2024 we got continuous monitoring of this star for 6 hours (one third of the orbital period of \ktwo), obtaining a total of 10 spectra with an exposure time of 1800\,s (two of them in-transit).
Additionally, we obtained three datapoints on 3-February-2024, 17-February-2024, and 18-February-2024. All spectra were obtained with the Fabry-P\'erot in fiber B to monitor the intra-night drift of the instrument. The data were reduced using standar procedures with the CARACAL\footnote{CARMENES Reduction And CALibration.} pipeline \citep{piskunov02,zechmeister14}, version 2.20.
We use our own developed cross-correlation algorithm, \texttt{SHAQ}, developed for the K-dwarfs Orbited By habitable Exoplanets experiment (KOBE, \citealt{lillo-box22}), to extract the radial velocities, correct for the drift velocities, and obtain CCF properties like FWHM and BIS. 
We obtain absolute radial velocities with an average RV uncertainty of 8 m/s for the 2018 campaign and 16 m/s for the 2024 campaign. However, these RVs vary by several km/s within each campaign and by more than 10 km/s between both campaigns. We will consider them as separate instruments for the RV analysis. The RVs and related indicators are shown in Table~\ref{tab:RVs}.

\subsection{MuSCAT3 multi-wavelength photometry}
\label{sec:muscat}
\begin{figure}
\centering
\includegraphics[width=0.50\textwidth]{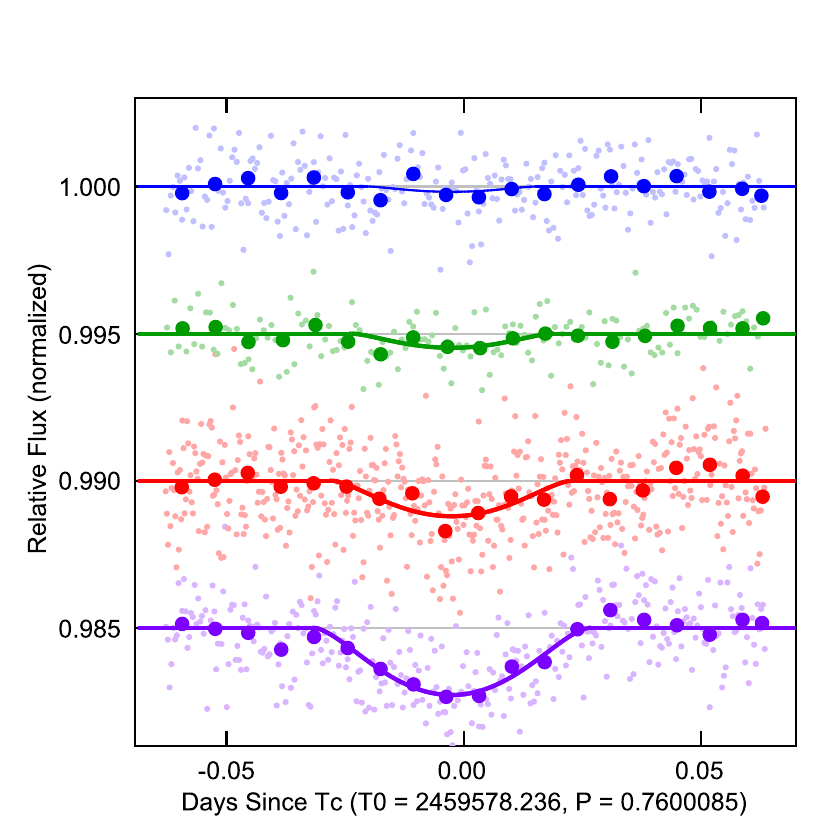}
\caption{LCOGT MuSCAT3 light curves of K2-399. The light curves from top to bottom are in the MuSCAT $g$, $r$, $i$, and $z_s$ bands. The small symbols show the unbinned data and the larger symbols show the same data in 10 minute bins. The transit model fits are overplotted. A V-shaped event was detected on-target with depths at mid-transit of 0.18, 0.48, 1.21, and 2.28 ppt in $g$, $r$, $i$, and $z_s$ bands, respectively, indicating that a fainter, much redder, star is blended in the $5.1$~arcsec follow-up photometric aperture and is hosting the eclipse.\label{fig:MuSCAT3}}
\end{figure}

We observed a full transit event window with the MuSCAT3 multi-band imager \citep{Narita:2020} of \ktwo{} on 30-March-2024 simultaneously in the four MuSCAT filters $g$, $r$, $i$, and $z_s$ (having bandpasses 400 - 550, 550 - 700, 700 - 820, and 820 - 920 nm, respectively) from the Las Cumbres Observatory Global Telescope \citep[LCOGT;][]{Brown:2013} 2\,m Faulkes Telescope North at Haleakala Observatory on Maui, Hawai'i. 
All images were calibrated by the standard LCOGT {\tt BANZAI} pipeline \citep{McCully:2018} and differential photometric data were extracted using {\tt AstroImageJ} \citep{Collins:2017}. We used circular photometric apertures of 5.1~arcsec, that excluded all of the flux from the nearest known neighbor in the \gaia{} DR3 catalog (\gaia{} DR3 3855957905329756416), which is $\sim47$~arcsec west of our target star (see Fig.~\ref{fig:tpf}). A V-shaped event was detected on-target with preliminary nominal depths at mid-transit of 0.18, 0.48, 1.21, and 2.28 ppt in $g$, $r$, $i$, and $z_s$ bands, respectively, indicating that a fainter, much redder, star is blended in the $5.1$~arcsec follow-up photometric aperture and is hosting the eclipse. The light curves with preliminary models overplotted are shown in Figure \ref{fig:MuSCAT3} and the data are available on {\tt ExoFOP}\footnote{\url{https://exofop.ipac.caltech.edu/tess/target.php?id=374200604}.}.

\subsection{Broad-band photometry}

We retrieved the observed broad-band photometry of \target{} from the Virtual Observatory SED Analizer (VOSA, \citealt{bayo08,bayo14}). In this process, we discarded several observations due to unknown uncertainties in their values. Table~\ref{tab:SED} shows the photometry used and the corresponding band passes. The spectral energy distribution (SED) based on this photometry is shown in Fig.~\ref{fig:SED}. 

\subsection{Spectroscopic stellar characterization}
\label{sec:StellarProp}

We used a combined HARPS spectrum for \target{} to estimate its stellar spectroscopic parameters ($T_{\mathrm{eff}}$, $\log g$, microturbulence, [Fe/H]) using the ARES+MOOG methodology described in detail in \cite{sousa21}, \cite{sousa14}, and \cite{santos13}. Details on this spectroscopic characterization are provided in Appendix~\ref{app:StellarProp}. The results of this analysis conclude that the dominant component of the spectrum in the visible range is a main-sequence G1 dwarf star with an effective temperature of  $5863 \pm 62$~K and a surface gravity of $\log{g}=4.06 \pm 0.11$ dex. We also estimate a turbulent velocity of $1.111 \pm 0.022$~km/s and a metallicity of $[$Fe/H$] = 0.335 \pm 0.014$ dex. Based on these parameters and using the calibrations from \cite{torres10b}, we obtain a mass of M$_{\star,A} = 1.31 \pm 0.03$~\Msun{} and a radius of R$_{\star,A} =  1.57 \pm 0.05$~\Rsun{}. These results are shown in Table~\ref{tab:StellarProp}.

\section{Evidence for demoting K2-399\,b as a planet}
\label{sec:analysis}

\subsection{Radial velocity modeling}
\label{sec:RVanalysis}

The radial velocity data obtained from the different instruments display large variations at a several km/s level. 
The generalised Lomb-Scargle (GLS, \citealt{zechmeister09}) periodogram of the dataset (accounting for the offsets resulting from the RV analysis described in this section) does not show any signal at the transiting period of 0.76 days (see top panel in Fig.~\ref{fig:gls} and the marked dotted red vertical line). Instead, the largest power signal in the GLS periodogram corresponds to a signal in the range 830 to 900 days. 

We model these radial velocities by using the standard approach already presented in other works of the same kind (see, e.g., \citealt{lillo-box20a}). In this case, we use a keplerian model parametrized by the orbital period ($P$), the time of conjunction ($T_0$), the eccentricity ($e$), the argument of periastron ($\omega$), and the RV semi-amplitude ($K$). We also add an instrumental offset ($\delta_i$) and a white noise term per instrument (jitter, $\sigma_i$).  

We used the \texttt{emcee} code \citep{emcee} to sample the posterior distribution of the parameters using a total of 44 walkers (four times the number of free parameters) and 50\,000 steps per walker in a first burn-in phase. We then focus on a small ball around the maximum a posteriori N-dimensional parameter space and run a second chain with the same number of walkers and half of the steps (i.e., 25\,000 steps). We checked for the convergence of the chains by  requiring that the length of the chain is at least 50 times the autocorrelation time as suggested in the \texttt{emcee} documentation. 

\begin{figure}
\centering
\includegraphics[width=0.48\textwidth{}]{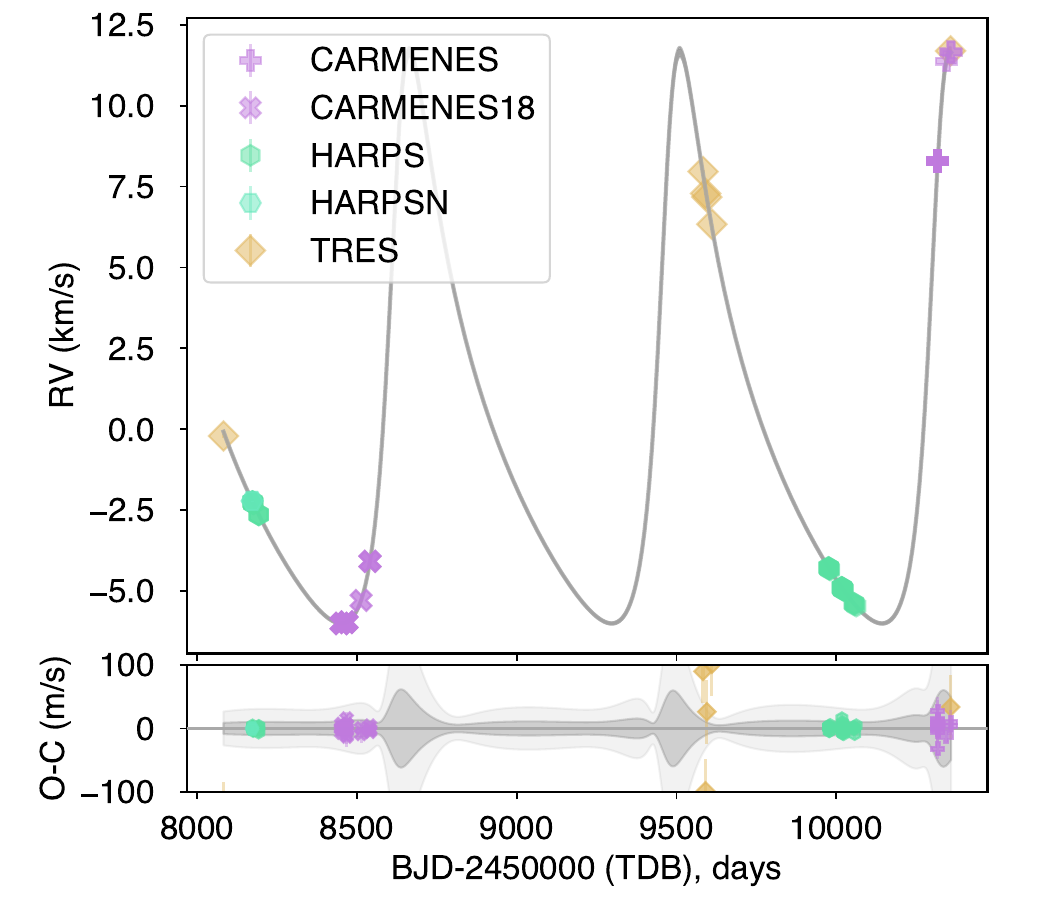}
\caption{Radial velocity time series (with the different instruments shown with different symbols and colors, see legend) and inferred radial velocity model corresponding to the 1-keplerian scenario. }
\label{fig:RVmodel}
\end{figure}

The confidence intervals of the marginalised distributions of the parameters as well as the prior distributions used are shown in Table~\ref{tab:posteriors}. The chains converge to a solution corresponding to a keplerian signal with a period of $P=846.62^{+0.22}_{-0.28}$ days and an RV semi-amplitude of $K=8.901^{+0.038}_{-0.050}$~km/s. The orbital architecture corresponds to an eccentric orbit of $e=0.4919\pm 0.0021$. These parameters correspond to a minimum mass of $0.4129\pm0.0066$~\Msun{} orbiting \targetA{} at such periodicity and with a highly eccentric orbital architecture. The data and median model are shown is  Fig.~\ref{fig:RVmodel}.

The overall root-mean-square (rms) of the residuals is 22~m/s, with no additional signals in the periodogram (see lower panel of Fig.~\ref{fig:gls}). In particular, no signal appears at the 0.76 days periodicity. Indeed, the rms per instrument is 5 m/s (HARPS), 13 m/s (CARMENES) and 36 m/s (TRES). The uncertainties on the individual measurements of the most precise instruments (HARPS and CARMENES) are at the same order as this rms, and the expected RV semi-amplitude of the transiting planet candidate (estimated by \citealt{christiansen22} based on empirical radius-mass relationships from \citealt{chen17}) was in the range 20-30 m/s. Our CARMENES dataset includes one night covering one third of this orbital period. If present around the main target, we should have detected its RV signal. Figure~\ref{fig:RVresiduals} shows the residual RVs phase-folded at the 0.76 days period after subtracting the long-period keplerian model, clearly showing no additional signals at the expected amplitude. 

Hence, we can clearly conclude that this dataset does not support the existence of a signal at 0.76-day periodicity around the star \targetA{} as previously claimed to correspond to an ultra-short period Saturn-like planet (\ktwo). Indeed, by adding a second keplerian signal to our model with tight priors on the period and time of conjunction from the transit signal, we can provide a maximum absolute mass for a transiting object around \targetA. By doing so, we can constrain the maximum RV amplitude at 0.76 days corresponding to $K_{\rm 0.76d} < 3.85 $~m/s (at 95\% confidence level). This would correspond to a maximum planet mass of $6$~\Mearth{}. This is however incompatible with a planet having an inferred Saturn-like radius of 6~\Rearth{} (as derived by \citealt{zink21,christiansen22}) whose expected mass is in the range $32 \pm 22$~\Mearth{} according to the same authors.

\subsection{Multi-band photometry modeling}
\label{sec:multiband}
We analysed the multi-band photometry of the transit event observed by MuSCAT3 (see Sect.~\ref{sec:muscat}). Details on the modeling of this dataset are provided in Appendix~\ref{app:muscat}. From such analysis, we find a clear chromatic effect, with the depth of the eclipse going from $2.43^{+0.18}_{-0.22}$ parts per thousand (ppt) in the redder $z_s$ band to being compatible with zero in the bluer $g$ band, see Figs.~\ref{fig:MuSCAT3} and \ref{fig:muscatFIT}. We conclude that this clear chromaticity implies a significant amount of blending light from a source unrelated to the eclipsing pair. This chromaticity and its color dependence is consistent with the previous RV analysis and shows clear evidence that the eclipses are not occurring on the G1 dwarf star (being the diluting source) but instead on the long-period companion. See Appendix~\ref{app:muscat} for details.

\section{New proposed scenario}
\label{sec:alternative}

The evidences for demoting \ktwo{} as a planet are clear from the analysis presented in Sect.~\ref{sec:analysis}. Here, we are curious to unveil the actual configuration of this system. 

So far, we know that \target{} has no chance-aligned companions as shown by the different high-spatial resolution images analysed in \cite{christiansen22} and accessible through the ExoFOP (at least to their sensitivity limits). Additionally, we know from the RV analysis in Sect.~\ref{sec:RVanalysis} that the star is accompanied by a long-period stellar companion with a minimum mass of 0.41~\Msun{}. The RV analysis also reveals no variations at the 0.76~days periodicity above 3.85 m/s at 95\% confidence.  From the multi-band photometry described in Sect.~\ref{sec:multiband} and Appendix~\ref{app:muscat}, we conclude that the transit presents a clear chromaticity, thus pointing to the long period low-mass companion as the host of the transit events (see, e.g., \citealt{parviainen19}). Hence, we have three components in the system: the bright G1 dwarf star dominating the spectrum as shown in Sect.~\ref{sec:StellarProp} (that we name $A$), a companion in a long-period orbit producing the large RV variations ($B$), and the object producing the eclipses in a short period ($C$). 

From the RV analysis and multi-band light curve analyses in Sect.~\ref{sec:RVanalysis}, the eclipses are produced on component $B$ and not in component $A$ (see also Fig.~\ref{fig:RVresiduals}). The question then is what are the properties of the eclipsing components. The most extreme scenarios that accomplish a mass distribution in agreement with the RVs ($M_B\sin{i}+M_C\sin{i}=0.41$~\Msun{}) are: i) most of the mass is in one of the components (thus $B$ being a $\sim$K7-M0 star\footnote{This accounts for some inclination different from 90$^{\circ}$.}) with the eclipsing object $C$ thus being either a grazing planet or a low-mass brown dwarf; or alternatively ii) the mass is equally split into both components ($M_B\sin{i}=M_C\sin{i}\approx 0.2$~\Msun), thus the eclipses being produced by a pair of mid-type M-dwarfs (e.g., M5+M5). From the evidences that we provide in Appendix~\ref{app:muscat} based on the multi-color analysis and in Appendix~\ref{app:SED} based on the SED analysis, we conclude that the second scenario represents the data significantly better ($\chi^2=374$ versus $\chi^2=1402$), composed of a G1 dwarf surrounded by a pair of similar-mass mid-type eclipsing M-dwarfs.

 However, in this scenario, assuming the period of 0.76~days, a secondary eclipse should have been detected, while it is not. This opens two alternatives: either 
1) the orbit is eccentric and it is oriented so that we only see the primary eclipse; or 
2) both components are of similar type (hence mass and radius), thus eclipsing each other and inducing same-depth eclipses but then with a period that is actually twice, i.e., $\sim$1.52~days.
In Appendix~\ref{app:NewPeriod} we describe the evidences we have in favor of the second scenario proving that the low-mass eclipsing binary accompanying \targetA{} has an orbital period twice that reported for \ktwo{} (see also Fig.~\ref{fig:TESStwice}). {This could also be solved by detecting the RV signal of the eclipsing binary. Indeed, an M5 star at this distance would contribute with a flux $10^{-2}-10^{-3}$ times the flux from the main G1 star in the near-infrared (NIR) regime (1-3 $\mu m$). With high SNR spectra in the near infrared, the RV signal of the eclipsing binary could be detected. However, our data in the NIR has an average SNR per pixel of 15, thus preventing any study in this regard.} 

Finally, we also note that recently developed validation tools applied to this system also strongly favor the HEB scenario against the planet hypothesis (see Appendix~\ref{app:AlternativeValidation} for additional information).

\section{Conclusions}
\label{sec:conclusions}

We have demonstrated with this work that the origin of the transit signals in \target{} is not of a planetary nature as previously validated by \cite{christiansen22}. Our RV data demonstrate the presence of a long-period component and the absence of the ultra-short period signal around the main target. The multi-color transit photometry shows the non-planetary origin of the eclipses, pointing to a low-mass binary, also suggested by the SED. This combined evidence points to a hierarchical triple system composed by an early G1-dwarf surrounded on a long-period ($\sim$847 days) orbit by a low-mass M-dwarf binary with a short period of $\sim$1.52 days (instead of the reported 0.76 days)  as the true scenario causing the transits mis-interpreted as of planetary origin. 

The previously confirmed planet \ktwo{} is hence demoted to a false positive hierarchical triple system.
The analysis presented here as well as in other previous works (e.g., \citealt{cabrera17}) demonstrates the clear need for a consensus on the definition of what we can consider as a confirmed planet. In this regard, we encourage the exoplanet community to {develop an} Exoplanet Confirmation Protocol to define commonly accepted (technique-independent) generic principles to establish the "Confirmed" status of a detected signal.

\begin{acknowledgements}
{We thank the referee, Alexandre Santerne, for his thorough revision of this manuscript that has improved its final quality.}
J.L.-B. is funded by the Spanish Ministry of Science and Universities (MICIU/AEI/10.13039/501100011033/) and NextGenerationEU/PRTR grants PID2019-107061GB-C61 and CNS2023-144309. A.C.-G. is funded by the Spanish Ministry of Science through MCIN/AEI/10.13039/501100011033 grant PID2019-107061GB-C61. B.M. is funded  by the former Spanish Ministry of Science and Innovation/State Agency of Research (MCIN/AEI/10.13039/501100011033/), grant PID2021-127289-NB-I00. D.G. gratefully acknowledge the financial support from the grant for internationalization (GAND\_GFI\_23\_01) provided by the University of Turin (Italy).
This research has made use of the SVO Filter Profile Service "Carlos Rodrigo", funded by MCIN/AEI/10.13039/501100011033/ through grant PID2020-112949GB-I00.
This research was funded in part by the UKRI, (Grants ST/X001121/1, EP/X027562/1).
We acknowledge a financial support from Astrobiology Center of NINS to obtain the observing time of CARMENES in 2019.
This project has received funding from the European Union's Horizon 2020 research and innovation programme under grant agreement No 730890 (OPTICON). This material reflects only the authors views and the Commission is not liable for any use that may be made of the information contained therein.
This research made use of the following software: \texttt{astropy}, (a community-developed core Python package for Astronomy, \citealt{astropy:2013,astropy:2018}), \texttt{SciPy} \citep{scipy}, \texttt{matplotlib} (a Python library for publication quality graphics \citealt{matplotlib}), \texttt{astroML} \citep{astroML}, \texttt{numpy} \citep{numpy}, and \texttt{AstroImageJ} \citep{Collins:2017}, TAPIR \citep{Jensen:2013}, 
\texttt{exoplanet} \citep{exoplanet:joss,
exoplanet:zenodo} and its dependencies \citep{exoplanet:agol20,
exoplanet:arviz, exoplanet:luger18,
exoplanet:pymc3, exoplanet:theano}.

This research has made use of NASA's Astrophysics Data System (ADS) Bibliographic Services, the SIMBAD database, operated at CDS, and the NASA Exoplanet Archive, which is operated by the California Institute of Technology, under contract with the National Aeronautics and Space Administration under the Exoplanet Exploration Program. We also used the data obtained from or tools provided by the portal \texttt{exoplanet.eu} of The Extrasolar Planets Encyclopaedia.
This research has made use of the Exoplanet Follow-up Observation Program (ExoFOP; DOI: 10.26134/ExoFOP5) website, which is operated by the California Institute of Technology, under contract with the National Aeronautics and Space Administration under the Exoplanet Exploration Program.
Funding for the TESS mission is provided by NASA's Science Mission Directorate. KAC acknowledges support from the TESS mission via subaward s3449 from MIT.
This work makes use of observations from the LCOGT network. This paper is based on observations made with the MuSCAT instruments, developed by the Astrobiology Center (ABC) in Japan, the University of Tokyo, and Las Cumbres Observatory (LCOGT). MuSCAT3 was developed with financial support by JSPS KAKENHI (JP18H05439) and JST PRESTO (JPMJPR1775), and is located at the Faulkes Telescope North on Maui, HI (USA), operated by LCOGT. MuSCAT4 was developed with financial support provided by the Heising-Simons Foundation (grant 2022-3611), JST grant number JPMJCR1761, and the ABC in Japan, and is located at the Faulkes Telescope South at Siding Spring Observatory (Australia), operated by LCOGT.
This work is partly supported by JSPS KAKENHI Grant Numbers JP24H00017, JP24K00689 and JSPS Bilateral Program Number JPJSBP120249910.
We are extremely grateful to the ESO and TNG staff members for their unique and superb support during the observations. 
\end{acknowledgements}

%
%

\bibliographystyle{aa} 
\bibliography{mybiblio} 

\begin{appendix}

\section{Ancillary analysis}

\subsection{Spectroscopic analysis}
\label{app:StellarProp}
The equivalent widths (EW) were consistently measured on the combined HARPS spectrum using the ARES code\footnote{The last version, ARES v2, can be downloaded at \url{https://github.com/sousasag/ARES}} \citep{sousa07,sousa15} for the list of lines presented in \citet[][]{sousa08}. The best set of spectroscopic parameters for each spectrum was found by using a minimization process to find the ionization and excitation equilibrium. This process makes use of a grid of Kurucz model atmospheres \citep{Kurucz-93} and the latest version of the radiative transfer code MOOG \citep{Sneden-73}. We also derived a more accurate trigonometric surface gravity using recent \gaia{} data following the same procedure as described in \citet[][]{sousa21} which provided a consistent value when compared with the spectroscopic surface gravity. 

The derived parameters from this study are shown in Table~\ref{tab:StellarProp}. As a double-check, we also estimated the fundamental stellar parameters by using the TRES spectra and the Stellar Parameter Classification (SPC). We also used the \texttt{param v1.3}\footnote{\url{http://stev.oapd.inaf.it/cgi-bin/param}} code for stellar parameter estimation \citep{daSilva06}. The results of this code after inputing the effective temperature and metallicity from the ARES+MOOG spectroscopic analysis, provide an age of $3.836 \pm 0.645$~Gyr, a mass and radius of $1.291 \pm 0.048$~\Msun{} and $1.689 \pm 0.093$~\Rsun{} and a surface gravity of $4.066 \pm 0.044$~dex. These values are compatible within the uncertainties with those from the empirical relations from \cite{torres10b}.

In all cases, the results are compatible with those from the co-added HARPS spectrum, which is used throughout the paper as our reference set of parameters.

\subsection{Spectral energy distribution}
\label{app:SED}

We analysed the broad-band photometry with the aim to constrain and test the hierarchical eclipsing binary (HEB) scenarios proposed for the configuration of the \target{} system. In practice, two scenarios have been proposed in this work for the eclipsing binary: a pair of similar mass mid-to-late M-dwarfs (labelled hereafter as G1+M5+M5) or a K7/M0 star plus a brown dwarf (labelled as G1+K7+BD). Given the slight differences in the spectral types, we investigated if this could be reflected in the SED. 

To this end, we built spectra based on the NextGen spectral models \citep{hauschildt99}. In order to weight the contribution of each stellar model to the composite SED for a given combination of objects, the stellar fluxes (which are given in the models in units of erg~cm$^{-2}$ s$^{-1} \AA^{-1}$) were multiplied by the corresponding radii squared. The parameters ($T_{\rm eff}$, $\log{g}$, $R_{\star}$) of the models were: G1 V (5863~K, 4.06~dex, 1.57~\Rsun), K7 V (4100~K, 4.65~dex, 0.63~\Rsun), M5 V (3060~K, 5.07~dex, 0.196~\Rsun). The parameters for the G1V component come from a spectroscopic determination from the HARPS spectrum (see Sect.~\ref{sec:StellarProp}); while the parameters for the K7V and M5V were taken from the table of stellar parameters by Mamajek\footnote{\url{https://www.pas.rochester.edu/~emamajek/EEM_dwarf_UBVIJHK_colors_Teff.txt}}; and for the brown dwarf the parameters were chosen to be 1000 K, $\log{g}\!=\!5.50$, $R\!=\!0.10$~\Rsun. A slight {reddening} $E(B\!-\!V)\!=\!0.036$ was applied to all models. The value was estimated by comparing the observed $B\!-\!V\!=\!0.658$ with the intrinsic colour of a G1 V star, 0.622.

\begin{figure}
\centering
\includegraphics[width=0.48\textwidth{}]{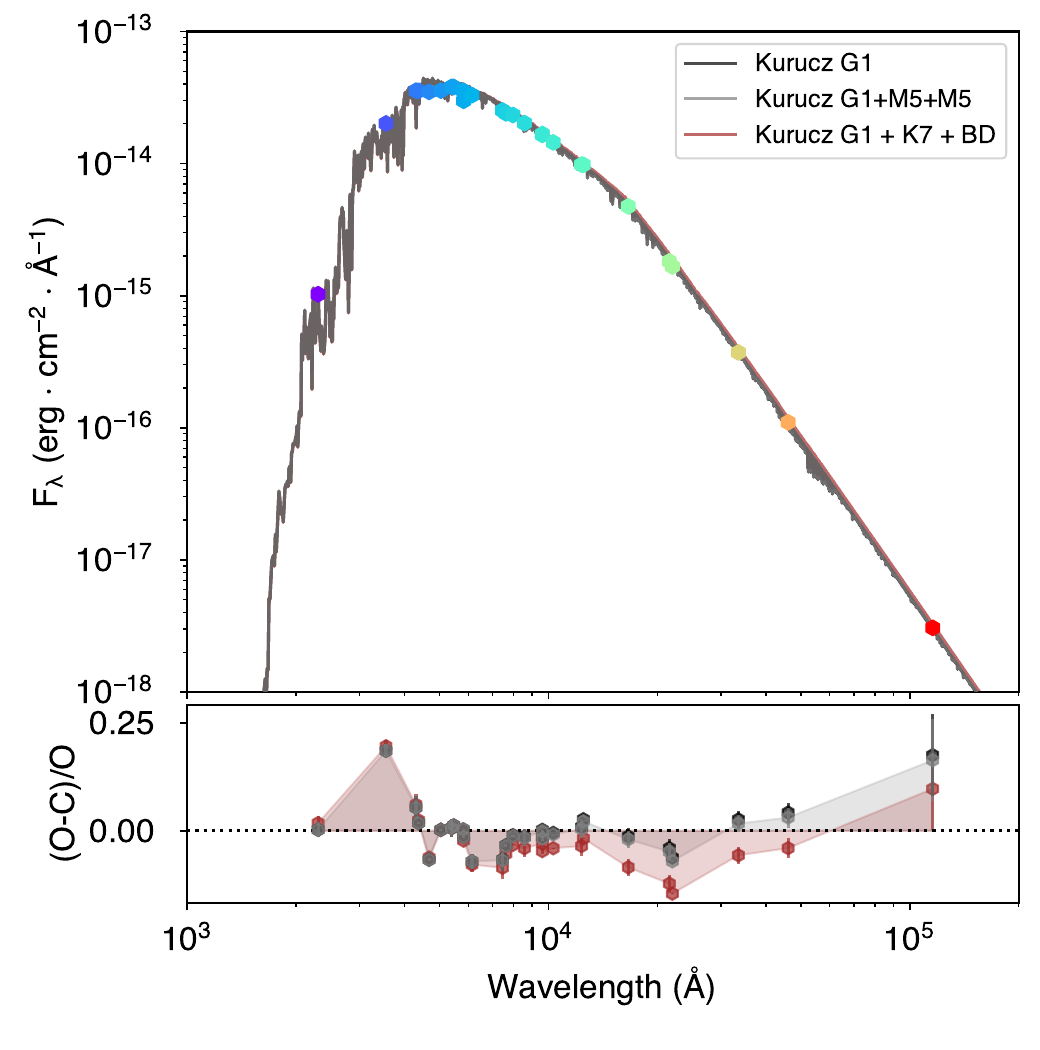}
\caption{Spectral energy distribution of \target{}. The observed fluxes are shown as colored symbols (coded by their wavelength) and the composite model of a G1 dwarf and two M5 dwarfs from the Kurucz sample of spectral models is shown in gray. The model has been normalized to the observed flux in the Ks band. }
\label{fig:SED}
\end{figure}

Figure~\ref{fig:SED} (top panel) shows the broad band photometry together with the two proposed models (G1+K7+BD and G1+M5+M5) also compared to a simple model of a single G1-dwarf. We scaled each model using the optical bands (filters with effective wavelength shorter than 7000~\AA) and determined the scaling factor through a least-squares methodology. To better visualize the differences between the models, we computed the synthetic photometry of the models at the observed bands by using the "Carlos Rodrigo" Filter Profile Service offered by the Spanish Virtual Observatory (\citealt{rodrigo12,rodrigo20,rodrigo24}). The lower panel in Fig.~\ref{fig:SED} shows the relative difference between the synthetic photometry from each model and the observed broad-band photometric measurements. As shown in this figure, the G1+K7+BD produces an infrared excess significantly larger than that observed in the data (red shaded region). By contrast, the G1+M5+M5 model is indistinguishable from the single G1 dwarf model and very much compatible with the observed data. Consequently, although the SED does not provide sufficient evidence to directly confirm the G1+M5+M5 scenario when compared to the G1 model based on the Occam's razor, it allows us to discard the G1+K7+BD configuration.

\subsection{Analysis of the multi-band transit photometry}
\label{app:muscat}

The strongly color-dependent transit (eclipse) depths from our MuSCAT3 data shown in Fig.~\ref{fig:MuSCAT3} point to a significant amount of blending light from a star unrelated to the eclipsing pair. To model this, we fit an eclipse model to all four bands simultaneously. Our model uses the same orbital and stellar parameters for the light curves in all four bands. The only differences are band-appropriate limb darkening, independent detrending for the four light curves, and a per-band free parameter that represents the amount of blending light in that filter, relative to the intrinsic flux from the eclipsing pair. By looking at how the required amount of diluting light varies with wavelength, we can infer some of the properties of the eclipsing system, especially since we know the stellar characteristics of the dominant star (see Sect.~\ref{app:StellarProp}).

\begin{figure*}
\centering
\includegraphics[width=0.99\textwidth{}]{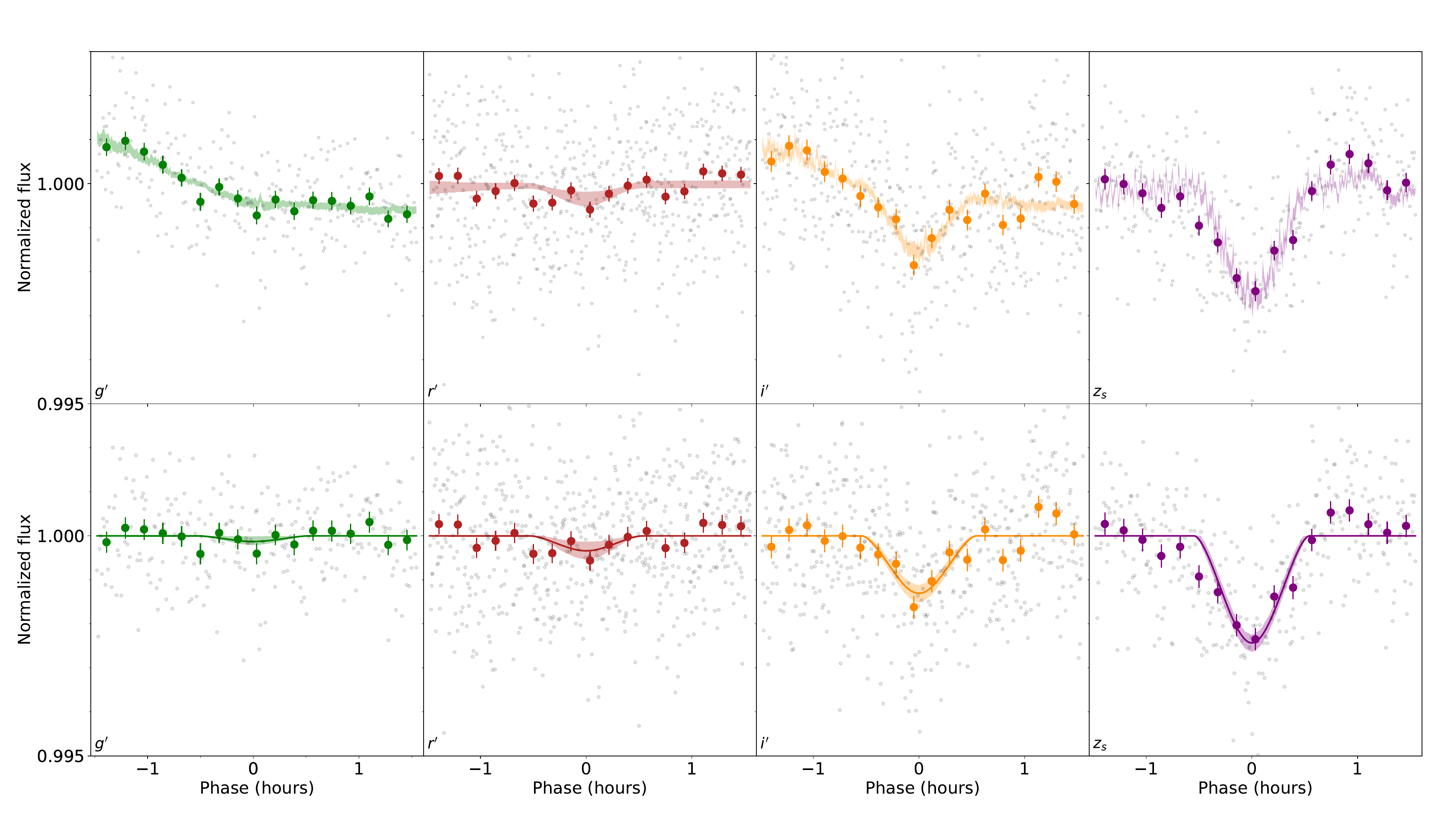}
\caption{Modeling of the MuSCAT3 light curves for all four bands, from left to right: $g$, $r$, $i$, $z_s$. Top panels show the raw light curves, while bottom panels show the detrended light curves and the corresponding model (all done in a simultaneous fit). The shaded areas around the fit lines are the 68.3\% credible interval in the MCMC.}
\label{fig:muscatFIT}
\end{figure*}

Based on the results from the RV analysis in Sect.~\ref{sec:RVanalysis}, we set a Gaussian prior on the stellar radius ratio with a mean of 1.0 and a standard deviation of 0.1. Imposing the assumption of nearly-equal-mass (and thus presumably equal effective temperature) stars allows us to also assume that the eclipse depth is intrinsically wavelength-independent, and to attribute the observed depth differences to the blending light. We build the model and sample the posterior of the parameters using the Python package \texttt{exoplanet} \citep{exoplanet:joss,
exoplanet:zenodo}, which uses Hamiltonian Monte Carlo with the No U-Turn Sampler \citep{Hoffman2014}, as implemented in \texttt{PyMC3} \citep{exoplanet:pymc3}. We ran seven independent chains with 1500 tuning steps and 3000 sampling steps each.  For every parameter of interest, the $\hat{R}$ statistic \citep{GelmanRubin1992} was less than 1.001, indicating convergence of the sampler.  The resulting models and detrended light curves are shown in Fig.~\ref{fig:muscatFIT}. 

\begin{figure}
\centering
\includegraphics[width=0.48\textwidth{}]{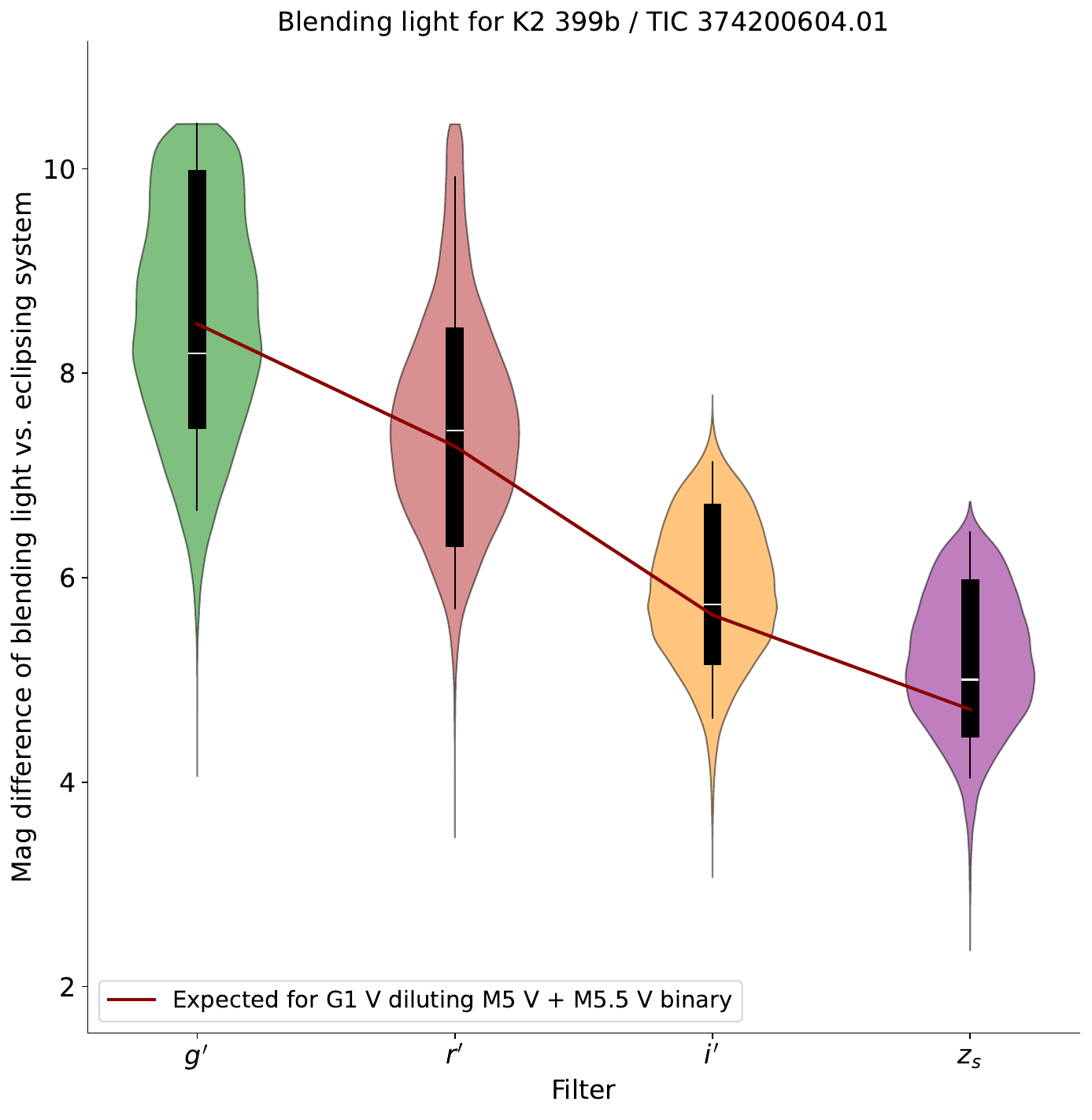}
\caption{Posterior distribution of the dilution factors (flux ratios between blending light and flux of the eclipsing system) for the four bands of the MuSCAT3 observations, expressed in magnitudes. The central black box shows the 68.3\% credible interval with the median marked as a white line, and the larger solid lines represent the 95.4\% credible interval. The surrounding color shows the posterior distributions from the MCMC run. The solid line shows the expected flux of a G1 V star in each band compared to a pair of M5 V stars (see text for details).}
\label{fig:dilution}
\end{figure}

Based on this analysis we can focus on the posterior distributions of the band-dependent dilution parameters. These results are shown in the violin plots in Fig.~\ref{fig:dilution}. Since the amount of blending light is specified relative to the continuum of the light curve, it essentially gives the flux ratio between the eclipsing system and the blend in that band. Hence, we overplot what we would expect if the blending light is from a G1V star, and the eclipsing system is an M5+M5.5. We use colors and absolute magnitudes from \citet{Covey2007AJ}, with a small correction from SDSS $z'$ to Pan-STARRS $z_s$ \citep{Tonry2012}. As shown in the figure, this scenario agrees well with the observations. We note that an M5+M5 model is also compatible, but with the $z_s$ band presenting a slightly larger difference ($\gtrapprox 1\sigma$).

In order to test the G1+K7+BD/planet scenario described in Appendix~\ref{app:NewPeriod}, we perform a similar analysis but in this case we restrict the radius ratio between the eclipsing components (i.e., $R_C/R_B$) to a maximum of 0.2. Although the results show that the data could still come from such model, the solution is quite marginal, with the model preferring both the $r^{\prime}$-$i^{\prime}$ and $i^{\prime}$-$z^{\prime}$ colors of the companion to be redder than a K7 or M0 star. Hence, based on the analysis of the multi-color eclipse MuSCAT3 data, the scenario composed of a G1 plus two similar-mass mid-type M-dwarfs seems to be preferred.

\subsection{Spectroscopic indicators}
\label{sec:indicators}

The HARPS data were also reduced with the HARPS-TERRA pipeline \citep{harps-terra}, which provides additional indicators based on specific spectral lines related to stellar activity like the S-index, H$\alpha$ or the sodium doublet lines. In this case, the average uncertainty of this dataset is 4.4 m/s with a dispersion in the uncertainties of 1.8 m/s. Given the larger consistency among the uncertainties from the DRS extraction, we use that RV time series in this paper (see Sect.~\ref{sec:HARPS}).

\begin{figure}
\centering
\includegraphics[width=0.48\textwidth{}]{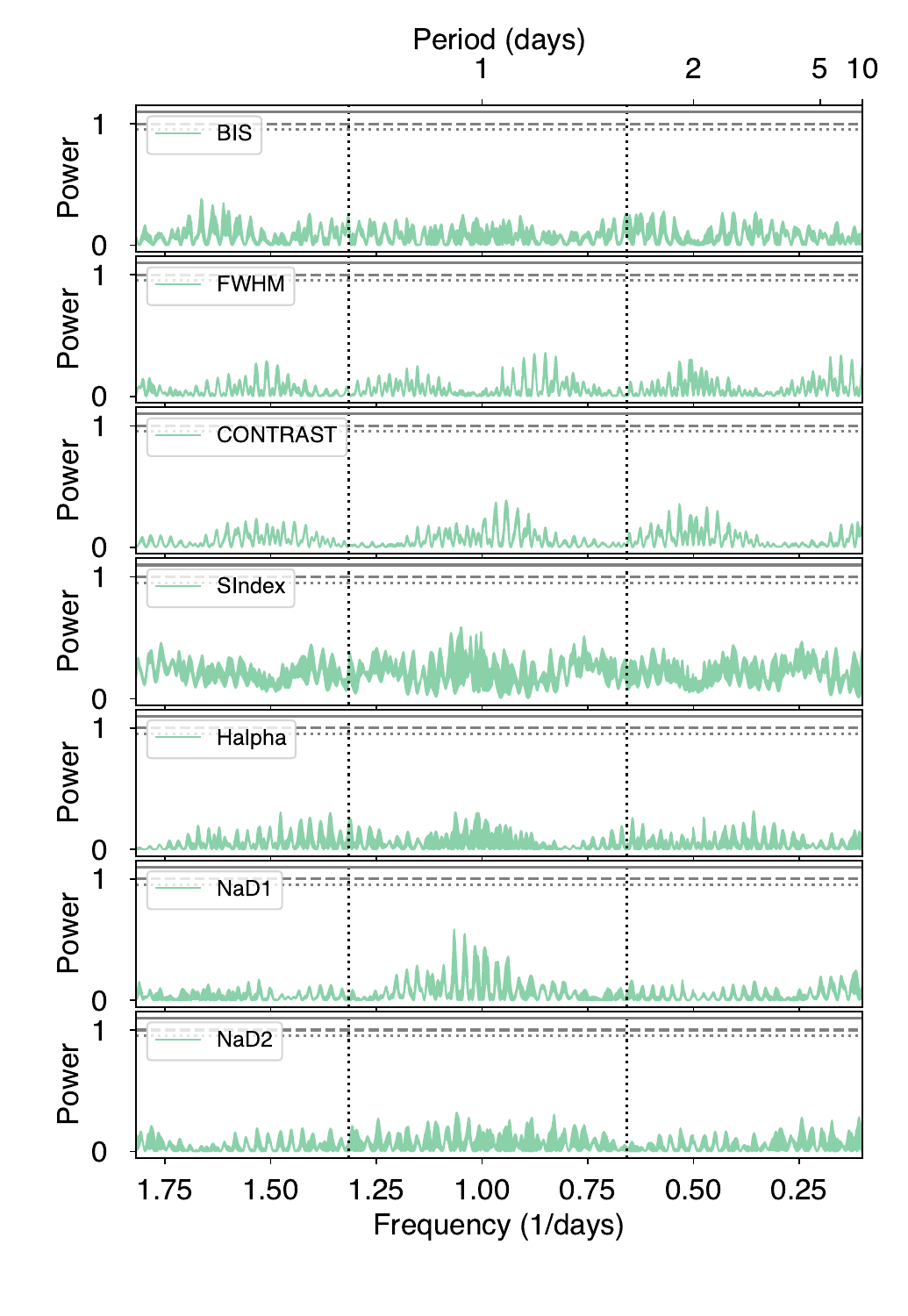}
\caption{Generalized Lomb-Scargle (GLS) periodogram of the activity indicators from the HARPS observations as determined from the HARPS-TERRA pipeline. The vertical dotted lines indicate the 0.76~days and twice this period (1.52~days). False alarm probabilities of 1\%, 5\%, and 10\% are indicated with horizontal dotted, dashed and solid lines, respectively. }
\label{fig:GLSindic}
\end{figure}

From the time series of the indicators from both the DRS and the HARPS-TERRA pipelines, we built the Generalized Lomb-Scargle (GLS) periodograms (\citealt{zechmeister09}) to look for correlations of these indicators with the periodicity of the eclipsing object. This is shown in Fig.~\ref{fig:GLSindic}, where we find no significant variations neither in the CCF asymmetry (BIS and FWHM) nor in the activity signal. Consequently, we can conclude that the star causing the effects is not eclipsing the main star \targetA{}.

\begin{figure}[H]
\centering
\includegraphics[width=0.48\textwidth{}]{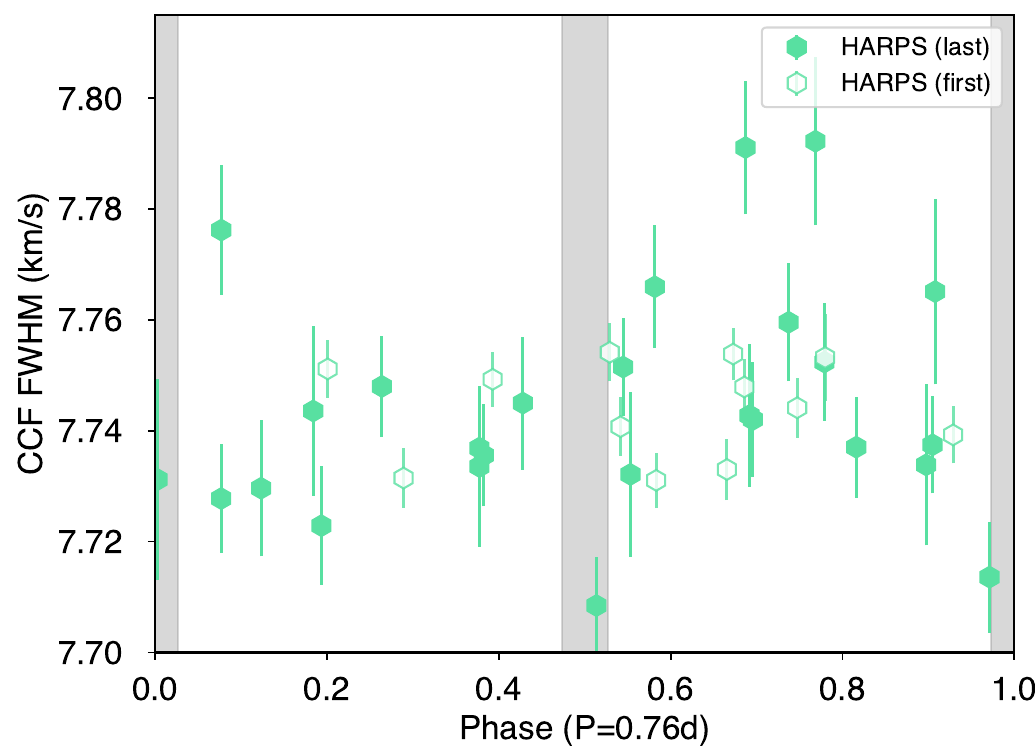}
\caption{Full-width at half-maximum (FWHM) of the cross-correlation function from the HARPS dataset phase folded with a 0.76-day period. Open symbols represent HARPS data obtained in the first campaign (in 2018), while filled symbols represent data from the last campaign (in 2023). The gray shaded regions show the location of the primary and secondary eclipses assuming a circular orbit for the binary.}
\label{fig:FWHMphase}
\end{figure}

However, given the complexity of the system, we explore the phase-folded diagram of the FWHM with the period of the binary. This is shown in Fig.~\ref{fig:FWHMphase}, where we have added a toy sinusoidal model as a dotted line to guide the eye. Even though the significance is still low, there seems to be a sinusoidal pattern in this diagram with an amplitude around 10 m/s. This small amplitude might be explained by an additional CCF corresponding to the combination of both $B$ and $C$ components. Interestingly, this dependency is enhanced when we only focus on the dataset from the 2023 campaign, when the binary was close to a quadrature of the orbit and hence the separation from the CCF of the main target is maximized. However in the 2018 dataset, the CCF of star $A$ and the CCF of $B$+$C$ are closer to conjunction and hence the signature in the FWHM is minimized. Indeed, if we focus on the filled symbols (the 2023 campaign), we see a larger phase dependency of the FWHM than in the opened symbols. This suggests that a second CCF coming from the combined $B$+$C$ binary is also somehow present in the spectra.  

Additionally, and more interestingly, one spectrum from HARPS (2023 campaign) was obtained during the eclipse of the binary (the symbol close to phase $\phi=1$ in Fig.~\ref{fig:FWHMphase}). This symbol is significantly below the expected value. This decrease in the FWHM of the global CCF of the system can easily be explained if we think of it as suppressing or at least fading the CCF corresponding to the $B$+$C$ binary as one of the components is being eclipsed and stops contributing to the global CCF itself, hence making the global CCF slightly sharper.

\begin{figure}[H]
\centering
\includegraphics[width=0.48\textwidth{}]{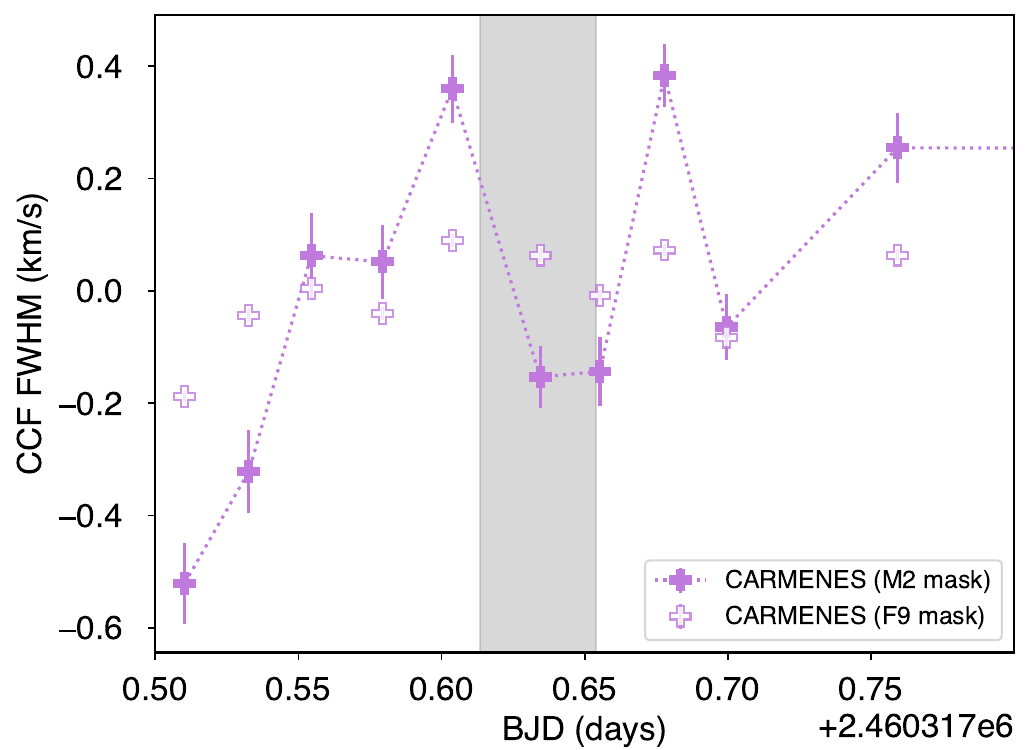}
\caption{Full-width at half-maximum (FWHM) of the cross-correlation function from the CARMENES time series for the night of 7-February-2024. Open symbols represent CARMENES CCFs obtained with the F9 mask (close to the spectral type of component $A$), while filled symbols represent CARMENES CCFs extracted by using an M2 mask (closer to the $B+C$ components). The dotted lines simply connect the datapoints from the M2 mask dataset to guide the eye. The gray shaded region shows expected time of the primary eclipse.}
\label{fig:FWHMphaseCARMENES}
\end{figure}

As a double check, we have also used the CARMENES data from the night that we observed the system during one of the eclipses. In this case, the observations were also performed close to a quadrature, and so we should see a decrease in the FWHM during the eclipse. Since the effect is very subtle, we perform another extraction of the CCF but now cross-correlating the spectra with an M2 mask instead of an F9. Figure~\ref{fig:FWHMphaseCARMENES} shows this analysis. Again, while in this case we cannot appreciate a smaller in-eclipse FWHM for the data from the CCF corresponding to the F9 mask (open symbols), it is very clear in the case of the M2 mask (filled symbols). Indeed, the decrease in the CCF FWHM is more abrupt in this case with the M2 mask than in the HARPS case. This is in fact what one would expect if the binaries are indeed of late spectral types, since the weight of the $B$+$C$ CCF against the CCF from the $A$ component is larger. 

However, despite all these indications, it is important to highlight that the significance of these signals is still too low to be considered as a clear evidence of the similar-mass scenario for $B$ and $C$. However, it adds support to the already discussed and favored scenario using other techniques.

\subsection{Eclipsing binary configuration}
\label{app:NewPeriod}

In Sect.~\ref{sec:alternative}, we proposed two scenarios to explain the absence of a secondary eclipse in the phase-folded diagram with a $\sim$0.76 day periodicity in the preferred G1+M5+M5 scenario. 

The first option (eccentric orbit of the B+C pair) allows any mass repartition among the B and C components. However, a 0.76-day period binary seems unlikely to have an orbit sufficiently eccentric as to prevent the secondary eclipse from occurring. Still, the large impact parameter keeps this possibility for eccentricities larger than around 0.15. However, for such a short period, these values seem unlikely. 

\begin{figure*}[t]
\centering
\includegraphics[width=0.99\textwidth{}]{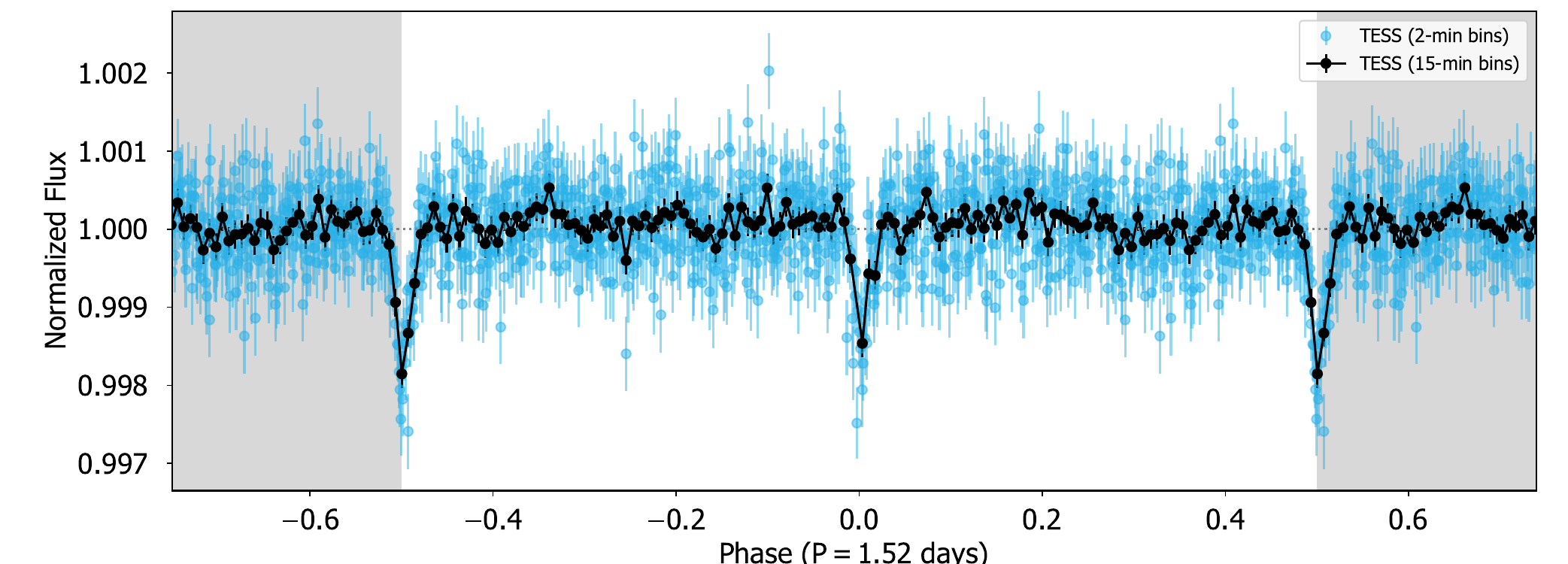}
\caption{TESS photometry folded with a period of 1.52 days (from the analysis of the multi-color band in Sect.~\ref{sec:multiband}), corresponding to twice the reported period from the K2 discovery in \cite{zink21}.Two different binnings are shown corresponding to equivalent sizes of 2 minutes (blue symbols) and 15 minutes (black symbols). A slightly different depth is apparent between phases $\phi=0.0$ and $\phi=0.5$, thus enhancing the similar-type low-mass star scenario for the eclipsing binary (see Sect.~\ref{sec:alternative}). }
\label{fig:TESStwice}
\end{figure*}

The second option (a twin pair of mid-M-dwarfs at twice the published period) is the most plausible from the data in hand. 
This alternative implies that the actual period is $\sim$1.52 days. Indeed, in Fig.~\ref{fig:TESStwice}, we plot the TESS photometry from the SPOC pipeline (using the PDC\_SAP flux) phase-folded with twice the reported period. As shown, there is also some evidence of differences in the depth of the odd-even eclipses (i.e., at phases $\phi=0.0$ and $\phi=0.5$ in the figure), {however it is still not significant enough ($180\pm135$~ppm) to be used as a claim .} 

We thus conclude that the most plausible scenario for this system is a hierarchical eclipsing binary where the brightest component ($A$) is a G1 dwarf, which is orbited in a 847~days period by a pair of mid-type M-dwarfs producing eclipses every 1.52 days. We note that under this scenario, the periodicity of the FWHM shown in Fig.~\ref{fig:FWHMphase} still holds because of the similar properties of both eclipsing components, making the width of the CCF to vary on a half-orbital-period rhythm. 

\subsection{Alternative validation techniques}
\label{app:AlternativeValidation}

An independent validation pipeline, known as \texttt{RAVEN}, is currently under development (Hadjigeorghiou et al in prep). \texttt{RAVEN} is an adaptation of the machine-learning based \textit{Kepler} validation tool presented in \citet{armstrong21} to the \textit{TESS} mission, and incorporates as part of the pipeline positional probabilities for true candidate host stars published in \citet{tpp}. We applied \texttt{RAVEN} to TOI-4838/K2-399 using \textit{TESS} lightcurve data supported by \gaia{} information, including nearby sources and RUWE values. The results are in strong agreement with the HEB scenario favoured by direct observations, finding a probability $>99.9\%$ that the candidate is a HEB in a direct Planet vs HEB test. Given that \texttt{RAVEN} is still under development we do not take this as evidence by itself, but note it in support of the scenario proposed from the above observations. It is also highly interesting that different validation pipelines on different datasets (here \textit{TESS} vs \textit{K2}) can give starkly different results, as noted by \citet{armstrong21}. This discrepancy is perhaps not surprising, particularly considering different datasets, but should be more clearly recognised when considering the status of validated planets in the context of the wider exoplanet population.

\subsection{Concerns on the validation process}
\label{app:validation}

Although we acknowledge the fact that this is a special case, it clearly evidences the need to properly establish the criteria for planet confirmation and signal verification. In this context, a community proposal is being discussed (the Exoplanet Confirmation Protocol, ECP) to decide upon the necessary criteria to consider a signal as a true confirmed planet. One of the key steps in this discussion is the verification of the origin of the signal, especially in the case of signals detected through the transit method that cannot be further followed-up with radial velocities due to a shallow RV signature or faint host star. In those cases, the community has opted for the so-called validation technique, based on discarding all other possibilities causing the light curve dimming that are not of planetary origin. 

In the case of \ktwo{}, the authors in \cite{christiansen22} used the \texttt{vespa} python module\footnote{As of 2023 (i.e., after the publication of the validation paper), the \texttt{vespa} algorithm was retired (and no longer maintained) in favor of \texttt{triceratops} \citep{giacalone22}, see \cite{vespa-retire}.}. This module computes the individual probabilities of four different non-planetary scenarios that could mimic the observed transit signal. These are the eclipsing binary (EB), the background eclipsing binary (BEB), the hierarchical eclipsing binary (HEB), and the case of a blended star hosting an actual planet with different properties (B1p). For the particular case of \ktwo, we have found that the actual scenario corresponds to a HEB. The probability of this scenario as reported in \cite{christiansen22} is $1.54 \times 10^{-4}$. Such a low probability of this scenario is unlikely to be accurately calculated given the evidence provided in this work. This points to either an underestimate of the capability of the K2 light curve to constrain of the transit shape, an underestimated a priorital probability for the HEB scenario in the validation software in the absence of additional follow-up data (e.g., multi-color photometry or RVs), or an error in consideration of the evidence from ancillary data during the validation process (e.g. the large RUWE or the slight difference in the odd/even eclipse depths). 

Overall, we highlight that published planets from validation processes, whether in bulk or individually, may still contain false positives. Indeed, if a 99\% probability threshold is used, naively we should expect 1 in 100 validated planets to be false. On this basis, we encourage the developers of such codes to pay particular attention to the a priorital probabilities and all available evidence. Additionally, testing validation results against as wide an array of alternatives as possible, whether different validation tools or independent follow-up, should be prioritised. Further, we encourage owners of exoplanet catalogues to distinguish validated planets from `confirmed' planets where possible.

\section{Figures}

\begin{figure}[H]
\centering
\includegraphics[width=0.48\textwidth{}]{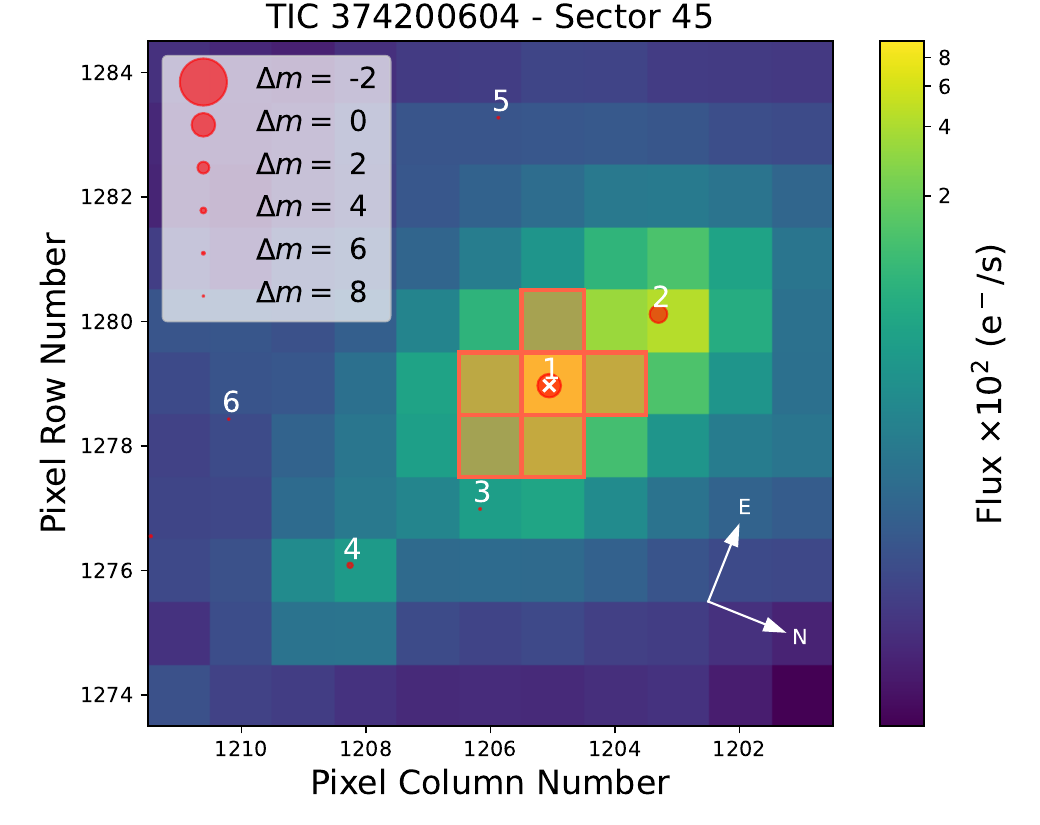}
\caption{Target Pixel File (TPF) of \target{} (also known as TIC\,374200604) from the TESS mission. The figure, created with the \texttt{tpfplotter} tool \citep{aller20}, shows the location of the target (white cross symbol) as well as the photometric aperture used by SPOC pipeline to extract the photometry (red shaded region) and the nearby \gaia{} sources (red circles with size scaled with the contrast magnitude against the main target).}
\label{fig:tpf}
\end{figure}
\FloatBarrier

\begin{figure}[H]
\centering
\includegraphics[width=0.48\textwidth{}]{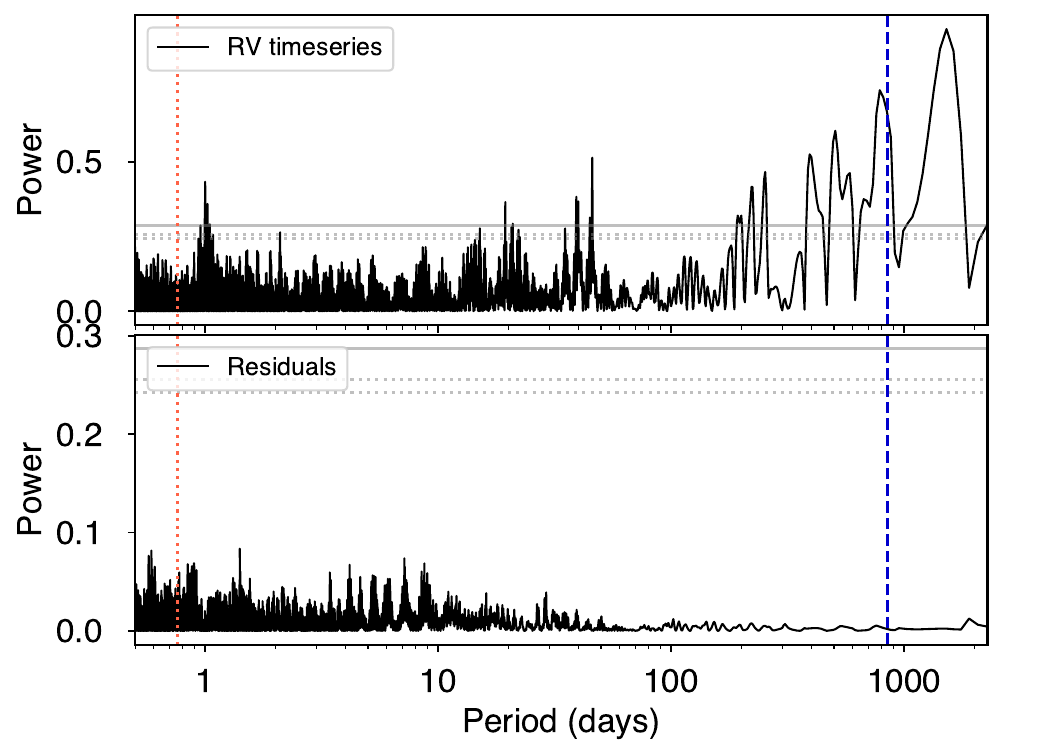}
\caption{\textbf{Top panel:} Generalised Lomb-Scargle (GLS) periodogram of the radial velocity time series from the TRES, HARPS and CARMENES instruments (assuming no instrumental offset). The vertical red dotted line indicates the reported transit period, while the vertical blue dashed line shows the converged periodicity of the RV data (see Sect.~\ref{sec:analysis}). \textbf{Bottom panel:} GLS of the residuals after subtracting the median RV model obtained in Sect.~\ref{sec:analysis}. In both panels, the horizontal dotted lines are the 1\% and 5\% false alarm probability (FAP) levels while the solid line corresponds to the 0.1\% FAP level. }
\label{fig:gls}
\end{figure}
\FloatBarrier

\begin{figure}[H]
\centering
\includegraphics[width=0.48\textwidth{}]{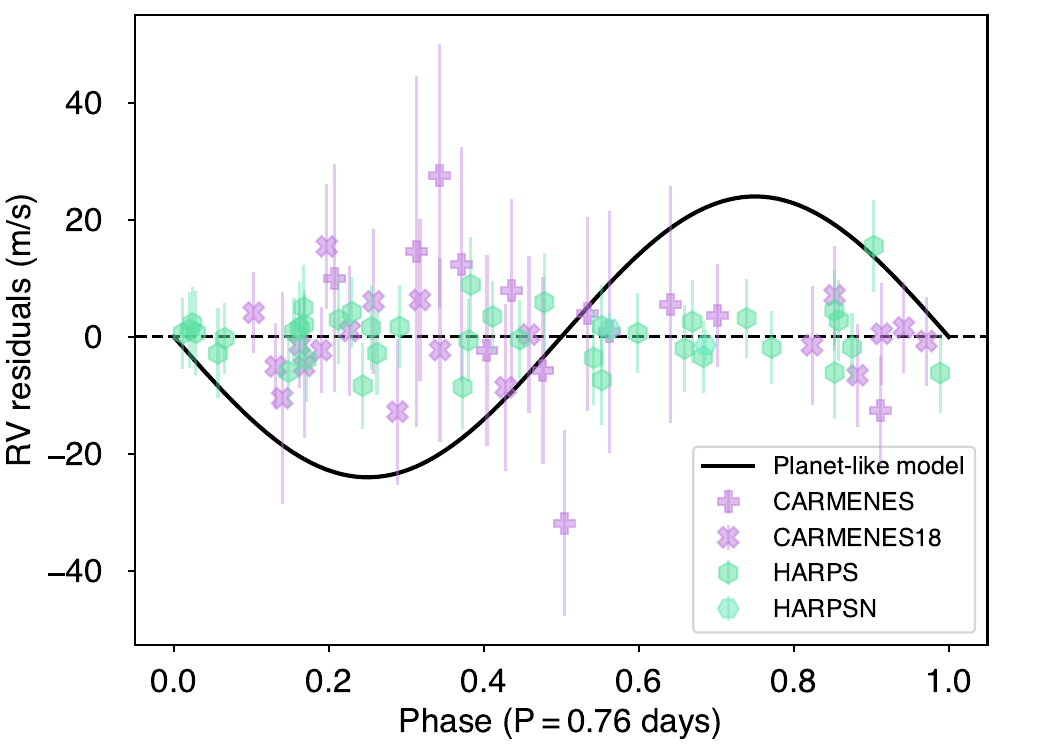}
\caption{Phase-folded CARMENES and HARPS radial velocities after subtracting the long-period keplerian model. The expected signal of the validated planet as estimated by \cite{christiansen22} is shown as a solid black line.}
\label{fig:RVresiduals}
\end{figure}
\FloatBarrier

\FloatBarrier

\section{Tables}

\begin{table}
 \setlength{\extrarowheight}{3pt}
\caption{Radial velocity measurements of \target. Only the first five rows are shown. This table is available in its full format at CDS. }
\label{tab:RVs}
\begin{tabular}{lccc}
\hline\hline
BJD  & RV  & $\sigma_{\rm RV}$ & Instrument  \\ 
(TDB) & (km/s) & (km/s) & \\ \hline 
2458172.66425265  & 25.5833    	   & 0.0059    & HARPS   \\
2458172.74227421  & 25.5840    	   & 0.0059	  & HARPS   \\
2458172.78903391  & 25.5852    	   & 0.0060	  & HARPS   \\
2458174.6535619   & 25.5385    	   & 0.0061	  & HARPS   \\
2458174.72069192  & 25.5386    	   & 0.0061	  & HARPS   \\
... \\
\hline\hline
\end{tabular}
\end{table}

\begin{table}
 \setlength{\extrarowheight}{3pt}
\caption{Broad-band photometry from \target{}}
\label{tab:SED}
\begin{tabular}{lcc}
\hline\hline
Band & Wavelength  & Flux$\times 10^{-14}$ \\ 
 & ($\AA$) & (erg $\cdot$ cm$^{-2}$ $\cdot$ s$^{-1}$ $\cdot$ $\AA^{-1}$) \\ \hline
GALEX.NUV & 2303.4 & $0.1026 \pm 0.0018$ \\
Johnson.U & 3551.1 & $2.013 \pm 0.031$ \\
APASS.B & 4299.2 & $3.562 \pm 0.092$ \\
Johnson.B & 4369.5 & $3.559 \pm 0.012$ \\
SDSS.g & 4671.8 & $3.491 \pm 0.013$ \\
GAIA3.Gbp & 5035.8 & $3.5710 \pm 0.0095$ \\
APASS.V & 5393.9 & $3.805 \pm 0.081$ \\
Johnson.V & 5467.6 & $3.799 \pm 0.011$ \\
ACS_WFC.F606W & 5809.3 & $3.5532 \pm 0.0051$ \\
GAIA3.G & 5822.4 & $2.9889 \pm 0.0076$ \\
SDSS.r & 6141.1 & $3.295 \pm 0.059$ \\
Johnson.R & 6695.8 & $3.2724 \pm 0.0068$ \\
SDSS.i & 7457.9 & $2.523 \pm 0.066$ \\
GAIA3.Grp & 7620.0 & $2.4036 \pm 0.0084$ \\
ACS_WFC.F814W & 7973.4 & $2.3314 \pm 0.0036$ \\
Johnson.I & 8568.9 & $2.3715 \pm 0.0039$ \\
GAIA3.Grvs & 8578.2 & $2.026 \pm 0.039$ \\
PS1.y & 9613.6 & $1.6760 \pm 0.0051$ \\
PS1.y & 9613.6 & $1.6481 \pm 0.0098$ \\
UKIDSS.Y & 10305.0 & $1.44498 \pm 0.00091$ \\
2MASS.J & 12350.0 & $0.982 \pm 0.022$ \\
UKIDSS.J & 12483.0 & $0.97965 \pm 0.00051$ \\
UKIDSS.H & 16313.0 & $0.42307 \pm 0.00027$ \\
2MASS.H & 16620.0 & $0.4745 \pm 0.0096$ \\
2MASS.Ks & 21590.0 & $0.1819 \pm 0.0039$ \\
UKIDSS.K & 22010.0 & $0.16633 \pm 0.00014$ \\
WISE.W1 & 33526.0 & $0.03719 \pm 0.00076$ \\
WISE.W2 & 46028.0 & $0.01099 \pm 0.00025$ \\
WISE.W3 & 115608.0 & $0.000306 \pm 0.000029$ \\
\hline\hline
\end{tabular}
\end{table}

\begin{table}
 \setlength{\extrarowheight}{3pt}
\caption{Spectroscopically derived stellar parameters of \targetA{} from both the HARPS co-added spectrum using the ARES+MOOG methodology and the TRES spectra using the SPC algorithm (see Sect.~\ref{sec:StellarProp})}
\label{tab:StellarProp}
\begin{tabular}{lcc}
\hline\hline
Parameter & ARES+MOOG  & SPC \\  \hline
T$_{\rm eff}$ (K) & $5863 \pm 62$ & $5726 \pm 130$ \\
$\log{g}$ (dex,cgs)    & $4.05 \pm 0.11$ & $4.12 \pm 0.15$ \\
$[Fe/H]$               & $0.335 \pm 0.014$  & $0.43 \pm 0.08$  \\
$v_{\rm turb}$ (km/s)  & $1.111\pm0.022$  & -  \\ 
$v\sin{i}$ (km/s)      &  - &  $5.5 \pm 1.2$  \\ 
M$_{\star,A}$ (\Msun)  & $1.31\pm0.03$ & - \\
R$_{\star,A}$ (\Rsun)  & $1.57\pm0.05$  & - \\
\hline\hline
\end{tabular}
\end{table}

\begin{table*}
\setlength{\extrarowheight}{5pt}
\centering
\caption{Prior and posteriors for the RV analysis of \target{} presented in Sect.~\ref{sec:analysis}.}
\label{tab:posteriors}
\begin{tabular}{lcc}
\hline\hline
Parameter  & Prior$^{\dagger}$ & Posterior \\ \hline 
Orbital period, $P_B$ [days] & $\mathcal{U}$(2,1000) & $846.62^{+0.22}_{-0.28}$ \\
Time of conjunction, $T_{\rm 0,B}-2400000$ [days] & $\mathcal{U}$(59551,60551) & $59684.31^{+0.72}_{-0.85}$ \\
RV semi-amplitude, $K_{\rm B}$ [m/s] & $\mathcal{U}$(0,100000) & $8901^{+38}_{-50}$ \\
Orbital eccentricity, $e_{\rm B}$ & $\mathcal{U}$(0,0.9) & $0.4919^{+0.0021}_{-0.0020}$ \\
Arg. periastron, $\omega_{\rm B}$ [deg.] & $\mathcal{U}$(-180,180) & $-49.59^{+0.23}_{-0.19}$ \\
$\delta_{\rm CARMENES}$ [km/s] & $\mathcal{T}$(27.7,0.3,27,28.5) & $27.756^{+0.066}_{-0.057}$ \\
$\delta_{\rm HARPS}$ [km/s] & $\mathcal{T}$(27.7,0.3,27,28.5) & $27.905^{+0.032}_{-0.042}$ \\
$\delta_{\rm HARPSN}$ [km/s] & $\mathcal{T}$(27.7,0.3,27,28.5) & $27.902^{+0.032}_{-0.041}$ \\
$\delta_{\rm TRES}$ [km/s] & $\mathcal{T}$(27.7,0.3,27,28.5) & $27.919^{+0.035}_{-0.033}$ \\
$\sigma_{\rm CARMENES}$ [m/s] & $\mathcal{U}$(-2.5,3.5) & $1.0^{+6.4}_{-5.6}$ \\
$\sigma_{\rm HARPS}$ [m/s] & $\mathcal{U}$(-2.5,3.5) & $0.5^{+4.2}_{-3.2}$ \\
$\sigma_{\rm HARPSN}$ [m/s] & $\mathcal{U}$(-2.5,3.5) & $0.9^{+8.3}_{-5.2}$ \\
$\sigma_{\rm TRES}$ [m/s] & $\mathcal{U}$(-2.5,4.0) & $8.8^{+4.9}_{-22}$ \\
\hline
\textit{Derived parameters} & & \\
\hline
Minimum mass, $m_{B}\sin{i_B}$ [\Msun{}] & (derived) & $0.4130\pm0.0066$ \\
Orbit semi-major axis, $a_{B}$ [AU] & (derived) & $1.916^{+0.014}_{-0.015}$ \\
Relative orbital separation, $a_{B}/R_{\star}$ & (derived) & $262.5^{+8.8}_{-8.3}$ \\
Stellar effective incident flux, $S_{B}$ [$S_{\oplus}$] & (derived) & $0.710^{+0.057}_{-0.054}$ \\
Stellar luminosity, $L_{\star}$ [$L_{\odot}$] & (derived) & $2.61^{+0.20}_{-0.19}$ \\
\hline\hline
\end{tabular}
\tablefoot{$^{\dagger}$Prior distributions are defined as: $\mathcal{U}(a,b)$ for a uniform distribution between $a$ and $b$, and $\mathcal{T}(\mu,\sigma,a,b)$ for a normal distribution with mean $\mu$ and standard deviation $\sigma$ constrained between values $a$ and $b$ (the so-called truncated Gaussian). }

\end{table*}

\begin{table*}
\setlength{\extrarowheight}{5pt}
\centering
\caption{Prior and posteriors for the multi-band photometry analysis of \target{} presented in Sect.~\ref{app:muscat}.}
\label{tab:phot-posteriors}
\begin{tabular}{lcc}
\hline\hline
Parameter  & Prior$^{\dagger}$ & Posterior \\ \hline 
Eclipsing system orbital period, $P_c$ [days] & $\mathcal{N}$(1.520017,0.0000002) & $1.5200154^{+0.0000010}_{-0.0000016}$ \\
Time of mid-transit, $T_{\rm 0,c}-2400000$ [days] & $\mathcal{N}$(59578.996,0.002) & $59578.995 \pm 0.001$ \\
Impact parameter, $b$ & $\mathcal{U}$(0,1.5) & $1.46^{+0.04}_{-0.48}$ \\
Radius ratio, $R_{\rm C}/R_{\rm B}$ & $\mathcal{T}$(1.0,0.1,0.5,2.0) & $1.00^{+0.09}_{-0.12}$ \\
\hline\hline
\end{tabular}
\tablefoot{$^{\dagger}$Notation for prior distributions is the same as used in Table \ref{tab:posteriors}, with the addition of $\mathcal{N}(\mu,\sigma)$ to denote a normal distribution with mean $\mu$ and standard deviation $\sigma$. }

\end{table*}

\end{appendix}

\end{document}